# QUADRUPLEX DETECTION IN HUMAN CELLS

David Monchaud, Ph.D.

Institut de Chimie Moléculaire de l'Université de Bourgogne

ICMUB, CNRS UMR 6302, UBFC Dijon, France

david.monchaud@cnrs.fr

**I. Introduction**.

"*You will never be able to hit a target that you cannot see*".[1] This wise advice suits perfectly the exploration of the world of the infinitely small, notably the word of nucleic acids. Nearly half a century of research was required to unveil the complicated processes the canonical DNA duplex is involved in. These heroic decades, which saw the rise of cross-cutting scientific disciplines (from molecular biology to chemical biology and epigenetics), have afforded invaluable insights into the many cellular processes in which the duplex participates.[2,3] They have provided critical clues about the cellular mechanisms underlying the very notion of life, and–because targets were made visible–as many novel therapeutic opportunities.

However, scarcely had the duplex revealed (not all) its secrets when alternative nucleic acid structures appeared. They differ from the duplex in many ways, the most distinguishing feature being their strandness (from two to four strands), which leads to the folding of higher-order architectures such as triplexes, DNA junctions and quadruplexes.[4-6] Understanding the dynamics of these new players was a novel, rather daunting challenge for the community of nucleic acids: how, where and when these structures fold in cells (if any)? From which sequences? With which stability? Etc. A new burst of questions that required a new portfolio of molecular tools aimed at characterizing furtive, certainly low-abundance structures in an environment (the nucleus) that was already difficult to characterize.

Efforts were thus invested to combine and use knowledge and know-how acquired during the study of duplex-DNA to assemble an array of molecular tools poised to address new and challenging issues. Among them, the detection of quadruplex in cells: do these four-stranded DNA structures exist in cells? Can they represent a novel class of therapeutically relevant targets? Etc. *"You will never be able to hit a target that you cannot see"*. This advice primarily emphasizes the need for tools able to cast light on quadruplexes in cells. In the following sections will be demonstrated how the creativity of scientists has been instrumental in the design of exquisite molecular flashlights that have been developed and are currently used for understanding how, where and when quadruplexes fold and unfold in cells.

**II. Four-stranded DNA structures and quadruplex ligands, a bit of history.**

The structure of the double DNA helix was unraveled in 1953 by Watson, Crick, Wilkins and Franklin.[7-9] The events that swirl around the elucidation of its molecular structure are among the most fascinating stories of modern science.[10,11] This discovery, which marked the start of modern genetics, drew on hard-won results acquired since the initial attempts by Miecher in 1871 to characterize the substance found in the nucleus, initially referred to as "nuclein" and now known as nucleic acids.[12] A part of these efforts was dedicated to the characterization of a given subclass of nucleic acid, the guanylic acid, initially isolated from bovine pancreas by Hammarsten in 1894. Levene (1909)[13,14] and Bang (1910)[15] noticed the tendency of concentrated guanylic acid solutions to form hydrogels; the exact nature of the structural unit responsible for the gelation process, the self-assembly of four guanines in a G-quartet,[16] was unveiled by Davies in 1962 by X-ray crystallography.[17] Inferred as early as in 1982,[18] the formation of G-quartets in biologically relevant guanine (G)-rich sequences (immunoglobulin switch regions, telomeres) was demonstrated *in vitro* by Gilbert (1988)[19] and Klug (1989)[20] *via* the study of sequences in which continuous guanine tracts assemble into G-quartets that self-stack to form higher-order G-quadruplex structures. Further structural support to this theory were soon provided *in vitro* by the first nuclear magnetic resonance (NMR, Feigon,[21] Patel[22] and Lilley, 1992)[23] and X-ray crystallography investigations (Rich, 1992[24] and Lilley, 1994).[25] As discussed in the following sections, the demonstration of the formation of quadruplexes in a cellular context has proved markedly more challenging.

Despite the lack of firm evidences of the existence of quadruplexes *in vivo*, chemical programs were soon launched to identify chemicals capable of interacting with quadruplexes, also referred to as G-quadruplex ligands. Neidle & Hurley reported in 1997 on the first prototype of a small molecule specially dedicated to interact with quadruplexes for therapeutic dividends,[26] giving dramatic impetus to the field of quadruplex ligands. This momentum has never slackened since then, with dozens of new ligands reported every year, displaying ever better quadruplex-interacting properties (affinity, selectivity).[27,28] Rapidly, the idea of using ligands endowed with specific spectroscopic properties (fluorescence) emerged as a way to detect quadruplexes in cells. The use of small-molecules is indeed a pledge of cell-permeability, the possible toxicity issue being easily tackled by antiproliferative investigations implemented prior to optical cellular imaging. First attempts were reported as soon as in 1999 with the use of metallic porphyrins on metaphase spreads from human cancer cells.[29] These first images raised more questions than answers, delineating the limits of this new research field to come: are these dyes quadruplex-specific enough (not to say specific for nucleic acids over other cellular components such as proteins)? Do their spectroscopic properties suited to detect low-abundance targets (mostly the amplitude of the fluorescence modulation upon interaction with quadruplexes only)? Etc. These first images were not convincing enough to build a consensus across scientific fields (chemistry, biophysics, biology, genetics) on the existence of quadruplexes in cells, and beyond this, on their relevance as antiproliferative targets. This skepticism compelled scientist to fine-tune their strategy and sharpened their molecular tools to provide undisputable proofs of the existence of quadruplexes in human cells. This two-decade quest, further described below, has led to the design of highly sophisticated molecular tools now able to deliver reliable evidences of the biological relevance of quadruplexes: antibodies on one side, a guarantee of high-affinity and specificity for quadruplexes, and fluorescent ligands on the other side, which offer bioavailability and tunable spectroscopic properties. The respective advantages and drawbacks of these two approaches (*e.g.*, poor bioavailability and technical access for the former, low quantum yields and off-target labeling for the latter) make them complementary techniques that are often implemented in concert.

**III. Immunodetection of quadruplexes in cells.**

Immunodetection is an invaluable technique for tagging cellular targets,[30] usually *via* a two-step protocol comprising the target labeling *per se* with a primary antibody followed by a signal amplification with a secondary antibody conjugated to a fluorescent reporter. This approach was successfully implemented for the detection of quadruplexes in cells; however, due to their intracellular location, quadruplexes were stained in fixed and permeabilized cells only, which might raise questions regarding their existence in functional, living cells.

The very first antibody reported to react with quadruplexes was *me$^V$IIB4*, described in 1998 by Hardin.[31] This antibody was not used for the immunofluorescent detection of quadruplexes in cells but was assessed *in vitro* for its ability to discriminate quadruplexes from other DNA structures including duplexes and triplexes. It was also shown to have a high affinity for the quadruplex sequences found in the ciliated protozoan *Stylonychia lemnae* telomeres d(TG$_4$) and d(T$_2$G$_4$).

Soon after, Plückthun reported on **Sty3** and **Sty49**,[32] also raised against the *Stylonychia lemnae* telomeric sequences d(TG$_4$T). **Sty3** displayed better affinity for parallel *versus* antiparallel quadruplexes, while **Sty49** was found to interact with both topologies. Only **Sty49** was found efficient to label quadruplexes in isolated *Stylonychia* macronuclei, after a two-step protocol using fluorescein isothiocyanate (FITC)-labeled secondary antibody. Collected images provided the first visual confirmation of the existence of quadruplexes in nuclei but indicated also that quadruplexes do form in vegetative nuclei only, *i.e.*, in absence of replication. This points towards a possible protective roles of quadruplexes at telomeric levels (capping), regulated by telomere end-binding proteins (TEBPs) that control quadruplex formation, as it was confirmed afterwards.[33]

Balasubramanian reported on the **hf2** antibody that was, similarly to *me$^V$IIB4*, described for its ability to interact with quadruplexes *in vitro*.[34] **hf2** was found to strongly discriminate between various quadruplex topologies (with a preference for c-kit quadruplex) and was subsequently used to pull-down quadruplex from genomic DNA, providing an unbiased, sequencing (seq)-based demonstration of the existence of quadruplexes in the double-stranded genomic DNA of the human breast cancer MCF-7 cells.[35] Regarding quadruplex immunodetection, the same group developed **BG4**, an antibody used to visualize both DNA and RNA quadruplexes in human cells (Figure 1a). The three-step protocol implemented to amplify the fluorescence detection (BG4 incubation *per se*, then secondary and tertiary labeled antibodies) allowed for visualizing quadruplexes not only in the nuclei of MCF-7 and osteosarcoma U2OS cells[36] and both the nuclei and cytoplasm of SV40-transformed human MRC5 fibroblasts,[37] but also in metaphase chromosome spreads (from human cervical cancer HeLa cells), suggesting that quadruplex formation occur during chromosome segregation. Interestingly, the pretreatment of cells with DNA quadruplex-specific (pyridostatin, **PDS**)[38] and RNA quadruplex-specific ligands (carboxyPDS, **cPDS**)[39] prior to cell fixation and **BG4** immunostaining increased the number of quadruplex *foci* (as compared to untreated cells), lending credence to both the existence of quadruplexes in human cells and their druggability with cell-permeable small-molecules. **BG4** was also used for tracking quadruplexes in human tissues:[40] the immunohistochemical labelling of stomach and liver cancer tissues revealed a higher density of quadruplex landscapes compared to non-cancerous tissues, supporting again the strategic relevance of quadruplex-directed anticancer strategy. Beyond quadruplex detection in cells and tissues, **BG4** was also instrumental for the development of G4 ChIP-Seq, a method that combines chromatin immunoprecipitation (ChIP) of quadruplexes and sequencing.[41] This protocol was implemented to map the genome-wide location of quadruplexes in epidermal keratinocyte HaCaT cells. G4 ChIP-seq confirmed the prevalence of the quadruplex landscape in a chromatin context (with approximately 10,000 quadruplex sites), proving beyond doubts their biological relevance.

Lansdorp reported on a series of immunofluorescence studies performed with the antibody **1H6** (Figure 1b). This antibody was found efficient to label both nuclei and metaphase chromosome spreads from HeLa cells,[42] and the macronuclei of *Stylonychia lemnae*.[43] Similarly to **BG4**, the pretreatment of cells with quadruplex ligands, **TMPyP4**[44] and **telomestatin**,[45] increased the number of quadruplex *foci*, further supporting the capability of **1H6** to detect quadruplexes in cells. However, it was subsequently reported that **1H6** cross-reacts with non-quadruplex targets, primarily single-stranded thymidines, highlighting the need to exercise caution when interpreting **1H6** sub-cellular binding patterns.[46]

Finally, **D1** antibody was also reported to be effective in immunostaining quadruplexes in human cervical cancer SiHa cells (Figure 1c),[47] and assessing the modulation of quadruplex landscapes upon ligand treatments, including N-methyl mesoporphyrin IX (**NMM**).[48]

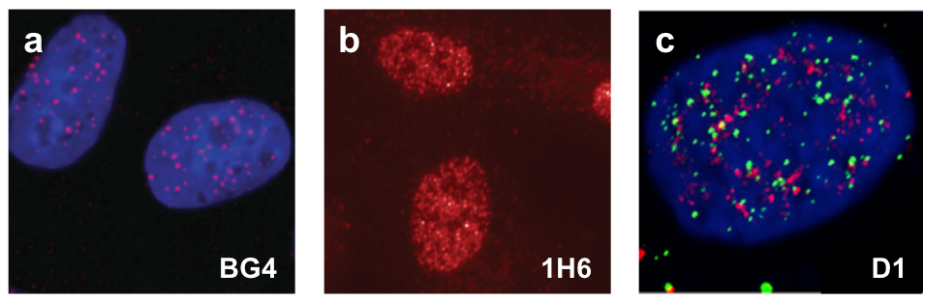

**Figure 1**. Immunodetection of DNA quadruplexes by (a) **BG4** in U2OS cells (red *foci*, the nuclei being counterstained with DAPI, in blue),[36] (b) **1H6** in HeLa cells,[42] and (c) **D1** in SiHa cells (green *foci*, co-incubated with an antibody raised against TRF2 (red *foci*), the nuclei being counterstained with DAPI, in blue).[47]

Altogether, these investigations have established that antibodies are exquisite molecular tools to visualize quadruplexes in cells by optical imaging. The only sour note of immunostaining is that investigations have to be performed with fixed and permeabilized cells, that is, with cells whose morphology is chemically altered, thus giving rise to doubts as to the biological relevance of observed staining patterns. This issue was alleviated thanks to the versatility of antibodies, which could be used in ChIP-seq protocols to gain reliable insights into the prevalence of the quadruplex landscapes in the human genome. The combination of these two approaches thus allowed the initial, *bona fide* doubts about the existence of quadruplexes in cells to be reliably dispelled. As further discussed below, a similar strategy was devised with small-molecules, combining chemical labelling and chemical affinity capture and sequencing (chem-seq)[49] to obtain complementary clues on the relevance of quadruplexes in functional, living cells.

### IV. Quadruplex-selective fluorescent dyes.

Quadruplex ligands are usually small-molecules (molecular weight < 2000 g.mol$^{-1}$) that display water-solubility and cell permeability properties suited to be used in living cells. The diversity of scaffolds that can be designed and synthesized through organic synthesis has offered access to a broad portfolio of molecular tools endowed with ever more accurate properties in a fully controllable manner. This explains the wide range of ligands now reported,[27,28,50,51] and made available for cell-based investigations. When displaying fluorescence properties, ligands become quadruplex-specific probes,[52-54] prone to be used in both fixed and living cells, the later allowing to track and detect quadruplexes in a functional context. These dyes could be divided in several categories as a function of their chemical nature (organometallic complexes *versus* organic molecules), cell-permeability (fixed cells versus live-cell imaging) and selectivity (DNA versus RNA quadruplexes, or both).

It is no coincidence that, historically, the first small-molecules described as quadruplex-interacting compounds, *i.e.*, a cyanine (**DODC**)[55] and a porphyrin (**NMM**)[56] in 1996, an anthraquinone in 1997,[26] a perylenetetracarboxylic diimide (**PIPER**)[57] in 1998 and a series of metal complexes of tetra(*N*-methyl-4-pyridyl) and tetra(*N*-methyl-3-quinolyl) porphines (**TMPyP4** and **QP3**, respectively)[29] in 1999, were also endowed with spectroscopic properties that make them suited to be used as quadruplex probes. These laid the foundation for a new area of research focused on the use of condensed polyaromatic molecules and metal complexes to track quadruplexes in cells. Several strategies were implemented over the past years, from the fine-tuning of the chemical structure of known quadruplex ligands (to make them fluorescent) to that of known dyes (to make them quadruplex-specific).

Most of the early examples of quadruplex ligands displayed spectroscopic (fluorescence) properties potentially suited to being used for tracking quadruplexes in cells. However, their polyaromatic nature along with that of their binding sites within the quadruplex structure (external G-quartets) makes them sensitive to the presence of their nucleic acid targets, which usually results in a "turn-off" (or "light-off")-type interactions, mostly *via* photoinduced electron transfer mechanisms.[58] While "turn-off" probes turned out to be useful for *in vitro* investigations,[52] they have only poor applicability for the detection of quadruplex in human cells. Astute strategies have been devised to counteract fluorescence quenching.[52,54,59,60] The fluorescence of properly designed aromatic molecules can be "turned-on" upon interaction with quadruplexes through a panel of possible mechanisms that include *i-* the restriction of internal rotation (RIR), which occurs with cyanine (*e.g.*, **DODC**,[55] thiazole orange (**TO**)[61] and Thioflavin T (**ThT**)),[62] carbazole (*e.g.*, **BMVC**)[63] or triphenylmethane dyes (*e.g.*, malachite green (**MG**),[64] crystal violet (**CV**));[65] *ii-* the protection provided by the DNA matrix against solvent-mediated non-radiative deexcitation (*e.g.*, porphyrins (e.g., **NMM**)[66] and metal complexes);[67] *iii-* the quadruplex-mediated disaggregation of non-emitting aggregates (*e.g.*, **NDI**)[68] or conversely, *iv-* the aggregation-induced emission (AIE, *e.g.*, the tetraphenylethene **TTAPE**);[69] *v*- the quadruplex-mediated molecular rearrangement of smart probes;[70] Etc. While hundreds of quadruplex-specific probes have been studied and reported, we will focus in the next sections on quadruplex-specific probes whose spectroscopic properties have been used to assess the existence of quadruplex landscapes in cells.

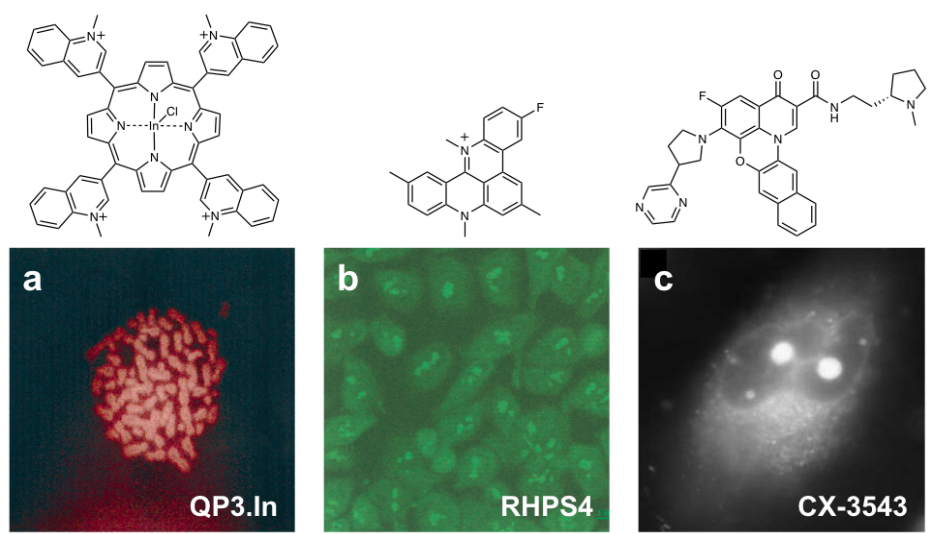

**Figure 2**. Images of (a) metaphase spreads of MCF7 cells live-treated by **QP3.In**,[29] and A459 cells live-treated with **RHPS4** (b)[71] and **CX-3543** (c).[72]

**IVa. Quadruplex-specific ligands with non-optimized spectroscopic properties.**
The first immediate strategy to track quadruplexes within cells is to exploit the fluorescence properties of established quadruplex ligands, even if these properties are non-optimized. As stated above, the first reported example reported (in 1999) relied on the live-incubation of MCF7 cells with a sub-toxic dose (40 µM, 3 days) the metal complex **QP3.In** prior to the preparation of metaphase spreads (Figure 2a).[29] The collected images were rather ill-defined and allowed for demonstrating the accumulation of the complex in the nuclei of cancer cells only. Soon after, Stevens reported on the synthesis and studies of a series of acridinium salts,[71,73] including **RHPS4** whose complex with a quadruplex was the very first to be elucidated by NMR.[74] They also exploited the intrinsic fluorescence properties of **RHPS4** to gain insights into the drug localization after short live-cell incubations (30 µM, 15 min) in both MCF7 and human lung carcinoma A459 cells by confocal microscopy (Figure 2b). Again, the low-resolution images did not allow for identifying precisely the quadruplex sites but they enlighten an accumulation in the nucleoli of A459 cells. Of note, higher-resolution images were collected recently in mouse embryonic fibroblast (MEF) cells using a high-speed confocal microscope, and showed an accumulation of RHPS4 in mitochondria.[75] Later on, the fluorescence of the fluoroquinolone **CX-3543** (or quarfloxin) was exploited for a more accurate analysis of the biodistribution of quadruplex ligands (5 µM, 1 h), confirming a privileged accumulation of quadruplex ligands in the nucleoli of A459 cells after live-cell incubation (Figure 2c).[72] This observation opened the way towards a redefinition of the in-cell targets of quadruplex ligands, which challenge the ribosomal biogenesis efficiently. However, these images failed in provided irrefutable arguments about the existence of quadruplexes in cells, and highlight the need of designing specially dedicated molecular tools to this end.

### IVb. From fluorescent dyes to quadruplex-specific probes

Another approach was thus devised with the aim of providing undebatable evidence about the relevance of quadruplexes in cellular contexts. This strategy was based on the use of–and then, the design of new systems from–commercially available nucleic acid stains. This approach was yet thoroughly followed but, as further detailed below, did not address all the aforementioned pitfalls.

### IVb.1. Fluorescent dyes with non-optimized quadruplex-interacting properties.

The illustrative example of a firmly established dye that found numerous applications as DNA staining agent is thiazole orange (**TO**). Its ability to interact with (owing to its positive charge) and label nucleic acids in a "turn-on" manner (RIR mechanism) whatever its secondary structure was thoroughly exploited, notably for the development of fluorescent intercalator displacement (FID) assays used to assess and quantify the apparent affinity of small molecules to duplexes[76] and/or quadruplexes.[77] However, **TO** does not discriminate between nucleic acid structures, making it of poor interest for the detection of quadruplexes in cells. Thioflavin T (**ThT**) is a closely related dye with better quadruplex-*versus*-duplex discriminating ability.[78] It was used for visualizing quadruplexes in polyacrylamide gels[79] and also found to be suited for cellular imaging (Figure 3a).[80] Collected images indicated a privileged accumulation of quadruplex-specific probes in the nucleoli of cancer cells (MCF7), both for post-fixation labelling (5 µM, 15 min) or *via* live-cell incubation protocols (5 µM, 24 h). These results indicated that **ThT** might interact with either ribosomal DNA (rDNA) or ribosomal RNA (rRNA), or both. The RNA-interacting properties of both **ThT** and the closely related analog **ThT-NE** were further investigated using the two probes for enlightening viral RNA (vRNA) in human hepatoma cells that express the full-length RNA genome of the hepatitis C virus (HCV).[81] Obtained images (Figure 3b) demonstrated a clear accumulation of the fluorescent *foci* within the cytoplasm of living cells (1 μM, 10 min), making this vRNA "turn-on" system a promising strategy for viral infection diagnosis. Another **ThT** analog, *N*-Isopropyl-2-(4-*N*,*N*-dimethylanilino)-6-methylbenzothiazole or **IMT**, was used for labelling quadruplexes in both fixed and living human cervical adenocarcinoma (HeLa) cells (Figure 3c).[82] **IMT** turned out to be particularly suited for live-cell investigations, allowing for monitoring the modifications of the DNA quadruplex landscapes real-time (4 μM, 0 - 90 min), notably upon treatment with the quadruplex ligands **PDS**, **RHPS4** and **TMPyP4**.[83]

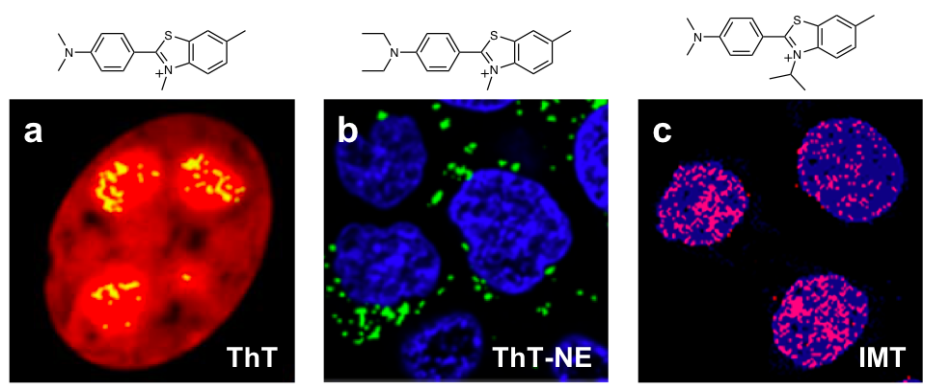

**Figure 3**. Images of (a) living MCF7 cells treated by **ThT** (yellow *foci*, the nuclei being counterstained with SYTO59, in red),[80] (b) living hepatoma cells infected by the hepatitis C virus treated by **ThT-NE** (green *foci*, the nuclei being counterstained with DAPI, in blue),[81] and (c) living HeLa cells treated by **IMT** (red *foci*, the nuclei being counterstained with DAPI, in blue).[82]

These results were interesting in that they showed that the use of commercial dyes (or closely related analogues) indeed enlighten what might correspond to quadruplex landscapes in cells in a straightforward manner. However, only additional manipulations (DNase/RNase treatments, co-incubation with firmly established quadruplex ligands, etc.) provide more circumstantial evidence of the *bona fide* relevance of the observed patterns. Also, using commercial dyes (or closely related analogues) leads to an *a posteriori* analysis of the fluorescent pattern rather than an *a priori* control over the dyes trajectory. Therefore, inspired by this wealth of promising data, numerous chemical programs have been launched to improve the performances of **TO**/**ThT**-based quadruplex probes.

## IVb.2. Fine-tuning the quadruplex-interacting properties of fluorescent probes

The two chemical motifs found in the **TO** scaffold, *i.e.*, the *N*-methylquinolinium and *N*-methylbenzothiazolium moieties (the latter being also found in the **ThT** scaffold), were truly inspirational and subsequently thoroughly exploited with the hope of obtaining quadruplex-specific probes with an alternative cellular activity. For instance, the π-surface of the **TO** was extended with a benzofuran to obtain the benzothiazole-fused benzofuroquinolinium dye **1**,[84] which proved useful for labelling living MCF7 cells (2 μM, 15 min), highlighting nuclei and nucleoli only (Figure 4a). Higher-resolution images were collected with a styryl-substituted **TO** referred to as **4a**,[85] which was useful for labelling both live and fixed human prostate cancer PC3 cells (5 μM, 15 min), the latter providing a clear visualization of a nucleolar accumulation of the dye (Figure 4b). The π-conjugation was further expanded introducing 3-vinylindole arms but the images of live PC3 cells incubated with compound **3c** (5 μM, 5 min) demonstrated, again, that the localization of *foci* remained globally centered on the nucleoli (Figure 4c),[86] without providing deeper insights into the localization of quadruplexes in cells.

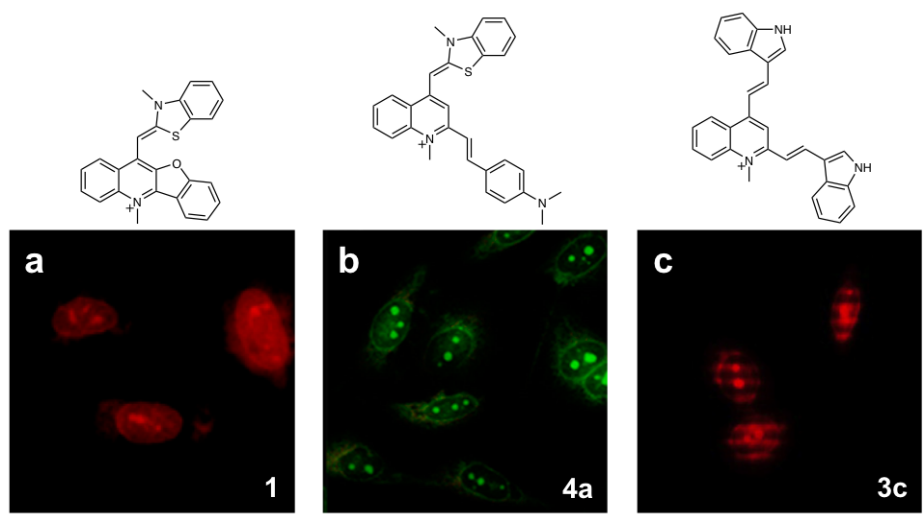

**Figure 4**. Nuclear labelling of (a) living MCF7 cells treated by the benzofuroquinolinium dye **1**,[84] and of living PC3 cells treated by the styryl-substituted thiazole orange **4a** (b)[85] and the vinylindole-substitued *N*-methylquinolinium **3c** (c).[86]

A higher level of precision was reached with *N*-methylbenzothiazolium-based probes. For instance, the cyanine **CyT** was found to be particularly sensitive to RNA quadruplexes and suited to the detection of RNA quadruplexes in living A459 cells (1.25 μM, 24 h), the foci being exclusively localized in the cytoplasm of the cells (Figure 5a).[87] Another benzofuroquinolinium dye named **CYTO** was implemented with both live and fixed PC3 cells (1 µM, 30 min) and leads to an interesting nuclear pattern from which nucleoli were found as the privileged labelling sites along with some–unfortunately unexploited–minor nuclear *foci* (Figure 5b).[88] The squaraine-based cyanine **CSTS** was found live cell-compatible (2 µM, 30 min) and accumulated in the lysosomes of MCF7 cells.[89] Similarly, another cyanine named **DMOTY** was found to accumulate in lysosomes of living HeLa and MCF7 cells (10 µM, 4 h) and be fluorescently responsive to the presence of quadruplexes (Figure 5c), making it a promising autophagy probes.[90]

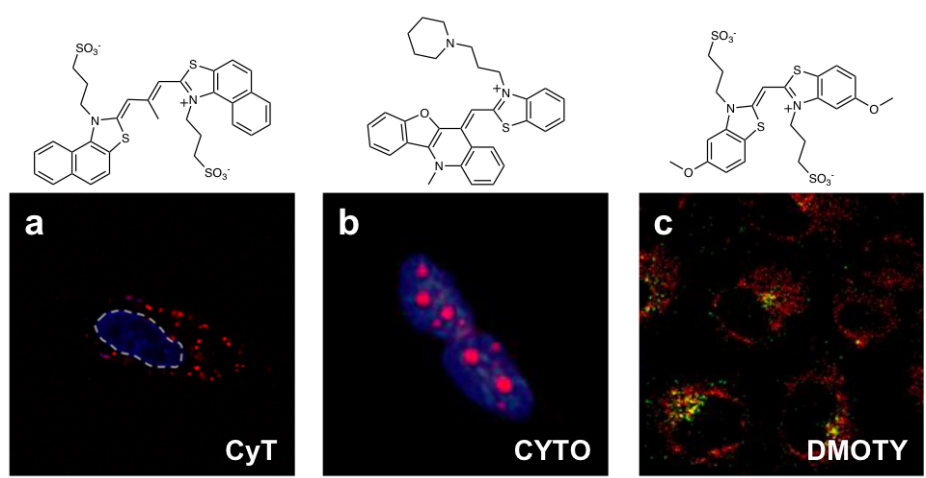

**Figure 5**. Labelling of (a) RNA quadruplexes by **CyT** in living A459 cells (red *foci*, the nuclei being counterstained with Hoechst 33258, in blue),[87] (b) the nucleoli of PC3 cells by **CYTO** (red *foci*, the nuclei being counterstained with DAPI, in blue),[88] and (c) the lysosomes of living HeLa cells by **DMOTY** (green *foci*, co-incubated with LysoTracker, in red).[90]

*N*-methylquinolinium-based probes were also further developed and studied. Two red-emitting dyes (with emission wavelength ($\lambda_{em}$) > 650 nm) were notably developed: **ISCH-1**, whose structure results in merged isaindigotone and coumarine scaffolds, was used with both fixed and living A459 and HeLa cells (5 µM, 30 min) and lead to an accumulation of foci in nucleoli (Figure 6a), presumably targeting rDNA quadruplexes as further demonstrated by ChIP assay;[91] **NCT**, whose structure results in conjugated **TO** and coumarine scaffolds, was used with living HeLa cells (1 µM, 15 min) and found to accumulate exclusively in mitochondrial DNA quadruplexes,[92] as demonstrated by co-localization with MitoTracker[TM] (Figure 6b). A very complete study was recently reported using the coumarine-based **QUMA-1** for labelling quadruplexes in both fixed and living HeLa cells.[93] The fixed conditions (1 µM, 15 min) were useful not only to show the very high selectivity of **QUMA-1** for RNA quadruplexes (cytoplasmic *foci* along with some isolated nucleolar sites, Figure 6c) but also to perform competitive experiments with RNA quadruplex-interacting ligands such as **cPDS**,[39] which results in the loss of **QUMA-1** *foci*. Live-cell conditions (0.5 µM, 3 h) uniquely provided insights into the dynamics of RNA folding and unfolding, notably upon addition of the helicase DHX36, which unfolds efficiently RNA quadruplexes resulting in the complete disappearance of the cytoplasmic *foci*.

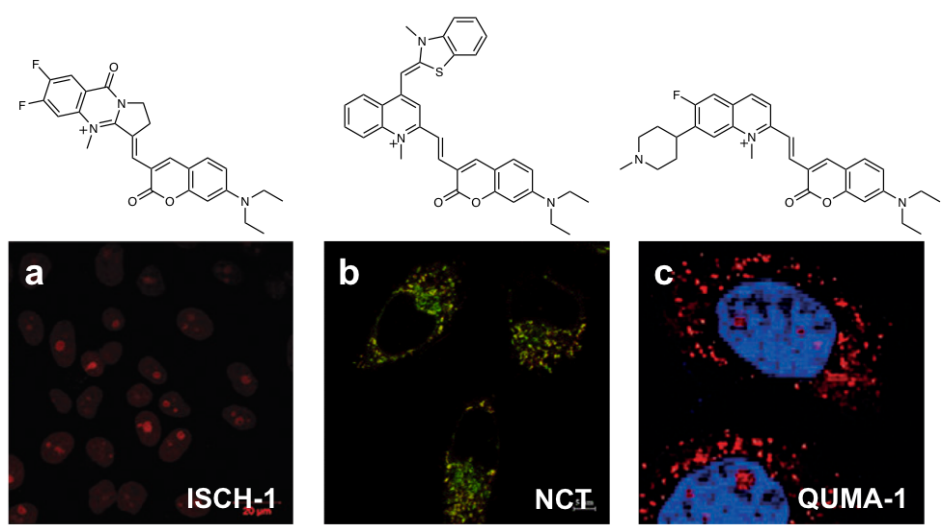

**Figure 6**. Confocal images of fixed HeLa cells treated by (a) **ISCH-1**,[91] (b) **NCT** (red foci, co-incubated with MitoTracker, in red, resulting in yellow *foci*),[92] and (c) **QUMA-1** (red *foci*, the nuclei being counterstained with DAPI, in blue).[93]

Tremendous results have thus been collected through these novel strategies. Fine-tuning the chemical structures of established fluorophores has indeed lead to quadruplex-specific systems that illuminated the versatility of the quadruplex landscapes in cells. However–and again–, additional manipulations (enzymatic digestions, co-incubation with ligands, helicases or specific antibodies, etc.) were still required to demonstrate the relevance of these approaches, and some pitfalls still exist, notably the influence of local microenvironments (pH, viscosity, etc.) that might trigger the fluorescence of the dyes in a quadruplex-independent manner. Smarter–and even more sophisticated–molecular systems were thus required to provide ever more convincing evidence about the existence of quadruplexes in cells.

**IVc. From quadruplex ligands to quadruplex-specific probes.**

Another strategy was thus implemented in the hope of providing molecular tools with first, a high affinity and selectivity for quadruplexes and then, the capacity of delivering readily detectable outputs once in interaction with their targets. This strategy was mostly based on the modifications of the chemical structure of well-established quadruplex ligands.

**IVc.1. Optimizing the scaffold of know quadruplex ligands**

As indicated above, fusing known quadruplex-interacting and **TO** scaffolds was soon envisioned as a way to design quadruplex probes suited to cellular investigations. This was pioneered fusing the **TO** scaffold with the pyridodicarboxamide (PDC) core of **360A** to provide **PDC-M-TO**,[94] with the acridine core of **BRACO-19** to provide a dual, pH sensitive and quadruplex-specific probe,[95] and with an isaindigotone core to provide **ISCH-1** (*vide supra*),[91] which was the only dye from that series that was attempted in cells. Some well-known nucleic acid stacking motifs were also revisited with successes, such as the quinacridone motif, notably with **QAB** that was made water soluble and bioavailable *via* the addition of solubilizing amine appendages,[96] the anthracene motif, notably with **9CI** that was used for labelling exogenous quadruplexes,[97] or the carbazole motif, whose quadruplex labelling properties have been thoroughly studied, spurred on by the compelling results obtained with **BMVC**[63] and its closely related analog *o*-**BMVC**. These two carbazoles were the first probes used for the detection of quadruplex in living human lung cancer CL1-0 cells (5 μM, 2 h), being suited to fluorescence lifetime imaging microscopy (FLIM) that enhanced the resolution of the collected images (Figure 7a).[98] They localized differently, **BMVC** being located in the nucleus with *o*-**BMVC** in the cytoplasm, likely in mitochondria. The carbazoles that belong to the **BMVC** family found numerous applications: they have been used to track exogeneous quadruplexes in CL1-0 cells (**BMVC**),[99] to illuminate mitochondrial DNA quadruplexes in living HeLa cells (**BMVC-12C-P**, 5 μM, 24 h; *o*-**BMVC-12C-P**, 5 μM, 4 h),[100] to implement a biosensing assay for the detection of cancer *versus* healthy cells (*o*-**BMVC**, 5 μM, 10 min, including fixed cancerous MCF-7 and CL1-0 cells *versus* non-cancerous fibroblast MRC5 and BJ cells, applicable in tissue biopsies as well),[101] and to assess the modifications of the quadruplex landscapes in HeLa and MRC-5 cells live-treated with the quadruplex ligands **BRACO-19** and **TMPyP4** (10 μM, 16 h) after fixation (*o*-**BMVC**, 5 μM, 10 min).[102] Many other derivatives have been synthesized and studied for live-cell imaging, such **BPBC** and **BTC**, found to localize in the cytoplasm and nucleoli of MCF7 cells (**BPBC**, 10 μM, 2 h),[103] and in the nucleoli of human liver cancer HepG2 cells (**BTC**, 5 μM, 4 h, Figure 7b).[104] Finally, another well-known nucleic acid stacking motif that was implemented for optical imaging purposes was the naphthalene diimide (NDI) unit. This motif was thoroughly exploited for the recognition of quadruplexes *in vitro*,[68] and examples were also reported for tracking them in cells. The spectroscopic properties of NDI, notably of core-extended NDI (**c-exNDI**), are based on a quadruplex-mediated disaggregation of non-emitting aggregates. Live incubation of human embryonic kidney HEK293T cells with **c-exNDI** (1 μM, 30 min) allowed for an accumulation of *foci* within the nuclei (Figure 7c), which partially overlapped with **1H6** immunostaining, and more precisely in the nucleoli, as demonstrated by nucleolin displacement investigations.[105]

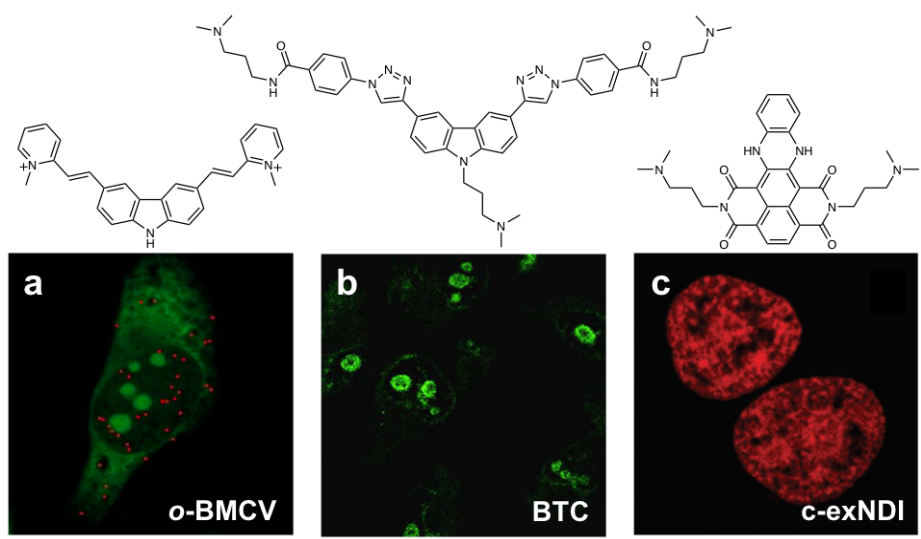

**Figure 7**. (a) Time-gated FLIM images of fixed HeLa cells treated with *o*-**BMVC** (red *foci* for decay times ≥2.4 ns, corresponding to quadruplexes, green *foci* for decay times <2.4 ns);[102] (b) nucleoli labelling of HepG2 cells by **BTC**;[104] (c) nuclear labelling of HEK293T cells by **c-exNDI**.[105]

### IVc.2. Adding tags to known quadruplex ligands

Attempts have been also made to covalently link a fluorescent tag directly to a quadruplex ligand scaffold. For instance, a BODIPY has been linked to a macrocyclic heptaoxazole (**L1BOD-7OTD**),[106] a **TO** and a BODIPY to the PDC core (**PDC-*L*-TO**[94] and **PDC-BODIPY**,[107] respectively), a Cy5 to a **PDS** derivative (**PDP-Cy5**)[108] and a fluorescein to a triarylimidazole (**IZFL-2**).[109] This approach turned out to be highly valuable for *in vitro* quadruplex detection: **L1BOD-7OTD** and **PDP-Cy5** were used to visualize quadruplexes by gel electrophoresis, while **PDC-*L*-TO**, **PDC-BODIPY** and **IZFL-2** were used to detect quadruplexes *via* fluorescence titrations. Doing so in cells represented another level of difficulty, given that the structure of the quadruplex ligand was drastically modified by the fluorescent appendage, diverting its intrinsic properties (pharmacodynamics, bioavailability) and target engagement. Despite these pitfalls, **PDP-Cy5** was successfully used for the detection of exogenous quadruplexes in HeLa cells (10 μM, 2 h, Figure 8a).[108]

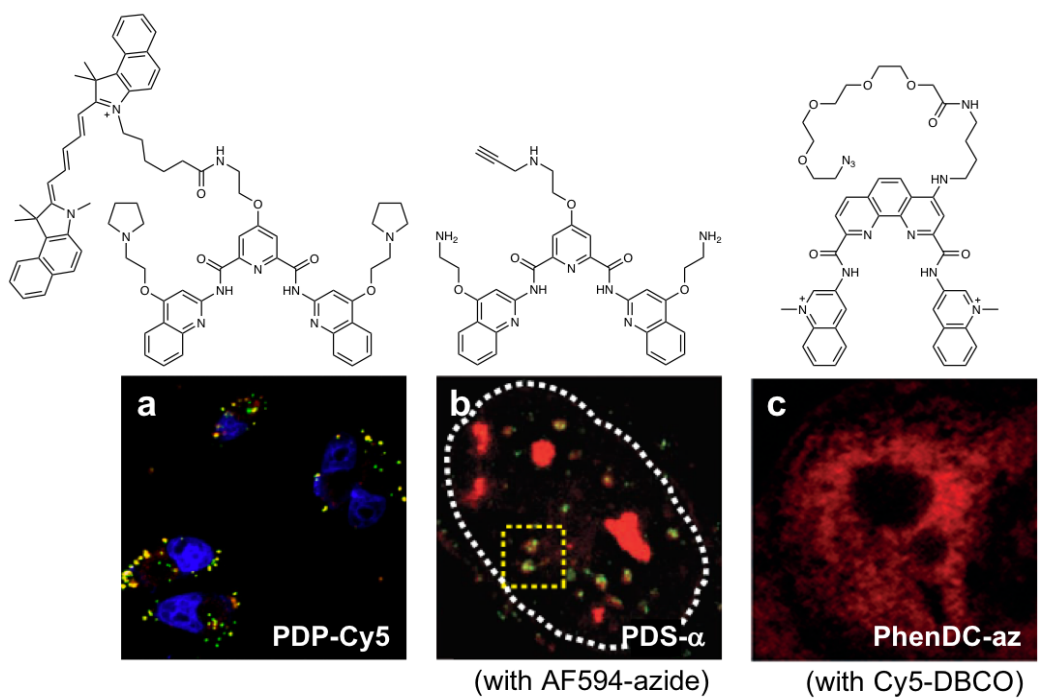

**Figure 8**. Detection of (a) exogenously added fluorescein-labelled quadruplex RNA in HeLa cells by **PDP-Cy5** (red foci, fluroscein in green, resulting in yellow foci, the nuclei being counterstained with DAPI, in blue),[108] (b) DNA quadruplexes in fixed U2OS cells thanks to **PDS-α** after CuAAC with AF-594 (red *foci*, co-localization with GFP-hPif1 in green, the nucleus being delineated by a dotted white line),[110] and (c) DNA quadruplexes in fixed HT29, after live-treatment with **PhenDC-az** and SPAAC with Cy5-DBCO.[111]

An elegant way to tackle these issues is to exploit the tremendous potential of bioorthogonal chemistry.[112] This was demonstrated using an alkyne-containing **PDS** derivative **PDS-α**,[110] which is minimally structurally modified as compared to the parent **PDS**, altering its biodistribution only marginally. The chemical labelling of **PDS-α** (1 μM, 12 h) once in its genomic sites within fixed human osteosarcoma U2OS cells using click chemistry with AlexaFluor594-azide allowed for a direct visualization of DNA quadruplexes (Figure 8b), as confirmed by co-localization experiments performed with the green fluorescent protein (GFP)-labelled quadruplex-specific helicase Pif1. This approach was extended with another well-established ligand, **PhenDC**,[113] using both copper-catalyzed azide-alkyne cycloaddition (CuAAC) between **PhenDC-alk** and Cy5-az, and strain-promoted azide-alkyne cycloaddition (SPAAC) between **PhenDC-az** and Cy5-DBCO (Figure 8c).[111] Experiments were performed live incubating human colon adenocarcinoma HT29 cells with clickable **PhenDC** derivatives (5 µM, 24h) prior to fixation and catalyzed or non-catalyzed AAC. The comparison of the two methods showed that caution must be exercised when interpreting the *in situ* click chemistry labelling results, due to possible relocation of intermediate metal complexes within the nucleoli.

### IVc.3. Specially dedicated small molecule probes

The least disruptive way to label a small molecule is isotopic labelling. This was demonstrated as early as in 2005 *via* studies performed with the PDC **360A**. The tritium-labeled **³H-360A** accumulated in the nuclei of living glioblastoma T98G cells (0.3 µM, 24 h) and the autoradiography of metaphase spreads of **³H-360A**-treated cells confirmed that genomic DNA was the target of the ligand (Figure 9a).[114] Similar experiments performed with T-lymphoblastic CEM1301 cells (characterized by longer telomere than T98G cells) and normal peripheral blood lymphocyte (PBL) cells clearly showed the preferential (but not exclusive) binding of **³H-360A** to the terminal region of the chromosomes, lending credence to the relevance of telomeric quadruplexes as one of the privileged targets for ligands. Another way not to disrupt the molecular organization of a quadruplex ligand to make it a fluorescent probe is to build the dye *ex nihilo*. This was demonstrated with NaphthoTASQ (or **N-TASQ**) whose design was based on a biomimetic interaction with quadruplexes. **N-TASQ** comprised four guanine arms that self-assemble into an intramolecular G-quartet upon interaction with their quadruplex targets. These arms are arranged around a fluorescent template (a naphthalene unit) whose fluorescence is "turned-on" by the formation of the template-assembled G-quartet (TASQ), making **N-TASQ** a twice-as-smart quadruplex ligand (both a smart ligand and a smart probe). **N-TASQ** was used with both fixed and living cells, interacting preferentially with RNA quadruplexes and accumulating in sub-cytoplasmic sites (stress granules and P-bodies) of living U2OS and MCF7 cells (5 µM, 48h, Figure 9b),[115] and labelling either DNA or RNA quadruplexes in fixed MCF7 cells, depending on the cell preparation protocol.[116] N-TASQ was also found useful to assess the the modulation of quadruplex landscapes upon ligand treatments, including **TMPyP4** and **BRACO-19**.[117] Another example of dedicated quadruplex probe is the triangulenium **DAOTA-M2**.[118] This probe turned out to be compatible for live-cell imaging performed with U2OS cells (20 μM, 24h) in which it bound to nuclear DNA (co-localization with Hoechst 33342 and DRAQ-5) and mitochondrial DNA (co-localization with rhodamine-123) but allowed for identifying quadruplexes *via* FLIM images (Figure 9c), given that its emissive lifetimes was strongly dependent on the topology of the nucleic acid targets (higher lifetimes (τ2 values) upon interaction with quadruplexes (up to 16 ns) *versus* duplex-DNA (<10 ns)).

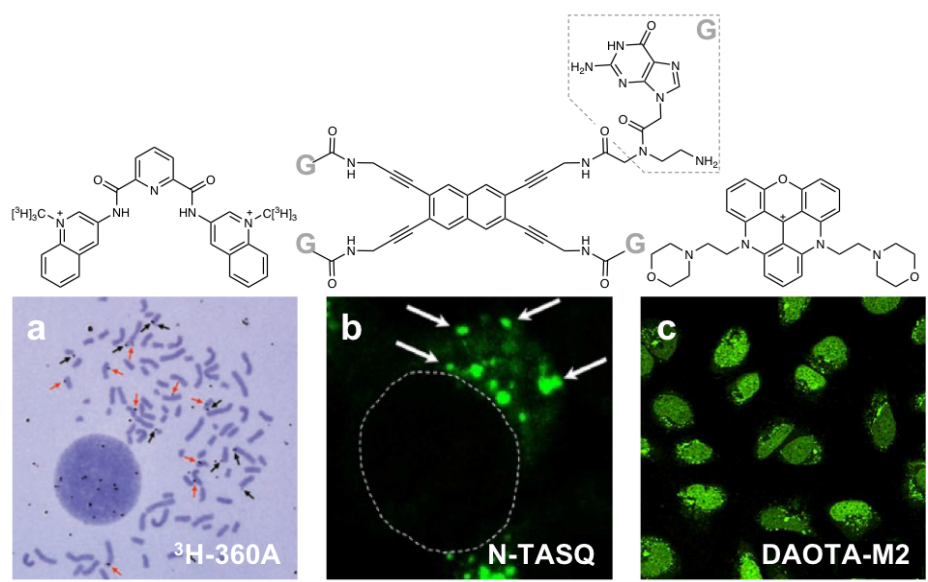

**Figure 9**. (a) Autoradiograph of metaphase spreads of T98G cells live-treated with $^3$H-360A (black arrows indicate chromosome ends, red arrows intrachromosomal sites);[114] (b) detection of RNA quadruplexes in the cytoplasm of MCF7 cells live-treated with **N-TASQ**;[116] (c) general staining obtained with U2OS cells treated with **DAOTA-M2**.[118]

### IVc.4. Using metallic probes to track quadruplexes in cells

After the initial impetus given by **QP3.In**, many metal complexes have been investigated as quadruplex-specific reporter probes. The quadruplex field was indeed a fertile playground for organic and inorganic chemists, owing to the manipulation of metal chelating moieties (sometimes referred to as ligands, avoided here for disambiguation) with broad aromatic surfaces (*e.g.*, phen, dppn, dppz, dpq, etc.),[119,120] prone to interact efficiently with the accessible quartet of quadruplexes. While dozens of metal complexes have been studied *in vitro*,[67] far less complexes have been used for cellular imaging, which can originate in the globally lower resolution of the obtained images as compared to those obtained with organic probes. The dinuclear ruthenium complex **[(phen)$_2$Ru(tpphz)Ru(phen)$_2$]** was for instance used to label the genomic DNA of both human MCF7 cells (500 μM, 1 h) and mouse lymphoma L5178Y-R cells (characterized by longer telomere than MCF7 cells; 200 μM, 30 min), but allowed for identifying quadruplexes *via* lambda stacking experiments (Figure 10a), given that its emissive maxima were dependent on the topology of the nucleic acid targets (630-640 nm for quadruplexes *versus* 670-700 nm for duplex).[121] The tetrakis-(diisopropylguanidinio)zinc phthalocyanine **Zn-DIGP** was used to label quadruplexes in living melanoma SK-Mel-28 cells (3 μM, 2 h)[122] or mouse embryo fibroblast NIH 3T3 cells (2 μM, 24 h, Figure 10b),[123] revealing foci localized in perinuclear regions, while the metal-free counterpart labelled nuclei only. Interestingly, the intrinsic chirality of the metal complexes was found to play an important role in the biodistribution of the probes. For instance, the two enantiomers of the ruthenium complex [Ru(bpy)$_2$(*p*-BEPIP)], referred to as **RM0627** were found to enter rapidly living breast cancer MDA-MB-231 cells (5 μM, 2 h) and accumulate either in perinuclear regions (the dextrorotary Δ-**RM0627**) or in the nuclei exclusively (the levorotary Λ-**RM0627**).[124]

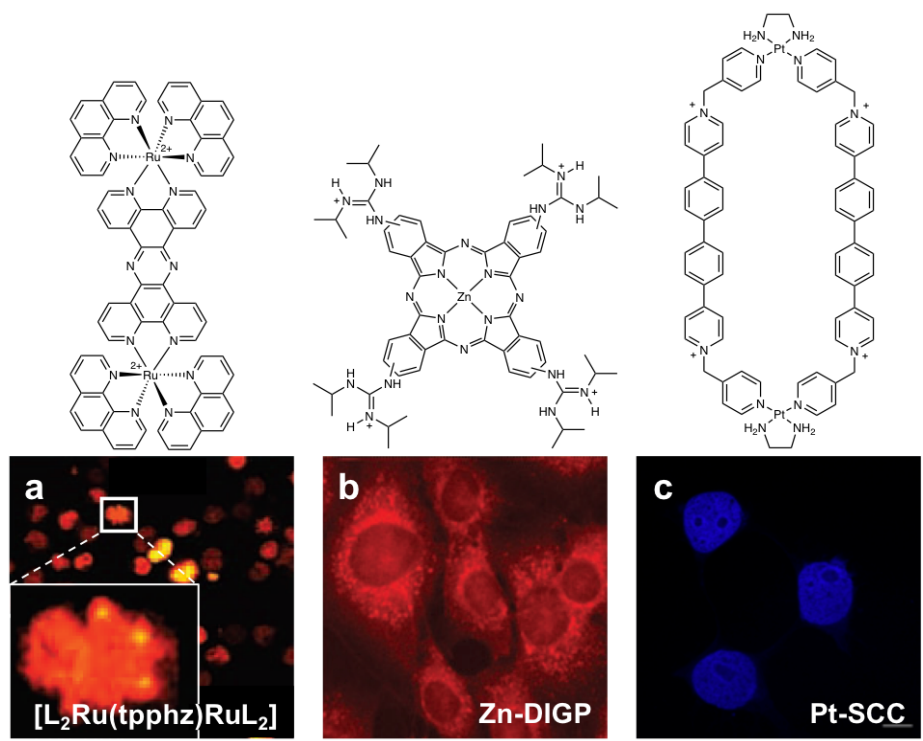

**Figure 10**. (a) Mutliple-emission images of L5178Y-R cells treated with **[(phen)2Ru(tpphz)Ru(phen)2]** (L = phen), highlighting the genomic DNA in red (emission: 670-700 nm) and the quadruplex *foci* (emission 630-640 nm) in yellow; confocal images of (b) NIN 3T3 cells treated with **Zn-DIGP**,[123] and (c) fixed MCF7 cells stained by **Pt-SCC**.[125]

More recently, the platinum complex **Pt-SCC** (for supramolecular coordination complex) was used in fixed MCF7 cells (50 μM, 30 min, Figure 10c) and lead to an intense nuclear labelling and the interaction with quadruplexes was confirmed *via* competition with quadruplex ligands (*e.g.*, **PDS**, **TMPyP4**) and co-localization with quadruplex-specific antibody (**BG4**).[125] Interestingly, **Pt-SCC** was found to provide a dual labeling, fluorescently staining the genomic DNA on one hand and increasing the contrast of nucleoli in direct transmission mode (DIC, for differential interference contrast) on the other hand, thus providing another modality (so called 'dark staining', relying on the absorption of light by the Pt-SCC/quadruplex complex) for the observation of quadruplex in cells.

Globally speaking, the optical images collected so far with metal complexes did not reach an exquisite level of resolution, which can be attributed to the fact that their structures, for most of them, relied on planar metal chelating moieties that failed in creating discriminating interactions with quadruplexes, as simple organic intercalators did. However, the chemical and structural diversity that uniquely offers this field of research, along with the various modalities of images that can be implemented with metal complexes, are the best promises for future advances and successes.

**V. Conclusion.**

The detection of quadruplex in human cells was and remains a challenging task. In spite of the massive efforts that have been invested for almost 20 years now, no turnkey solutions have been delivered and the examples discussed above highlighted that only multipronged approaches (co-localization of small molecule probes and antibodies; competition between established ligands and probes; etc.) can provide solid evidence about their existence in cells. The visualization of quadruplexes involves probes (antibodies, chemicals) suited to optical imaging but the very notion of detection of quadruplexes is boarder: it also encompasses the sequencing-based techniques that have been developed recently to gain insights into the relevance and prevalence of the quadruplex landscapes.[126-128] Once again, the role of quadruplex-specific antibodies and chemicals was instrumental in these investigations: for instance, the antibody **BG4** was the central molecular tool for the development of G4 ChIP-seq (*vide supra*),[41] which assessed the formation of DNA quadruplexes in the human genome; the small molecule **PDS** was used in both G4-seq[129] and rG4-seq[130] to promote and/or stabilize

DNA and RNA quadruplexes, respectively, to make them detectable during the sequencing analysis steps; and biotinylated **PDS**[131] and TASQ-based **BioTASQ**[132] was useful to purify (pull-down) and identify quadruplexes (*i.e.*, G4RP-seq).[133] These examples demonstrate again that (bio)molecular tools created to detect quadruplexes have to be versatile to be used in complementary fields, since only cross-disciplinary investigations can provide reliable insights into the cellular relevance and functions of quadruplexes. These examples have also casted a bright light on antibodies and small molecules but scientists have other strings in their bow, such as fluorescently labelled proteins (*e.g.*, the helicase GFP-hPif1, *vide supra*) or short oligonucleotides (*e.g.*, the oligonucleotide-templated reactions (OTR)[134] or G-quadruplex-triggered fluorogenic hybridization (GTFH)[135] approaches). This highlights the need for the quadruplex community to both keep growing and pull together to develop transdisciplinary research programs aimed at providing ever more convincing evidence to support therapeutic strategies based on the quadruplex targeting with *ad hoc* drugs.

"*You will never be able to hit a target that you cannot see*". We are now able to hit this target. The examples presented above showed that scientists have now a panel of molecular lights to enlighten quadruplexes in our genome and transcriptome, a target that can no longer hide–and will be hit–even in the remotest parts of the human cells.